\documentstyle{aipproc}

\begin{document}

\title{
Quark Mixing Possibilities in Extensions of the Standard Model and the 
Electromagnetic Current
}

\author{
A. Garc\'{\i}a$^a$,\
R. Huerta$^b$,\
and
G. S\'anchez-Col\'on$^{b,}$\footnote{Also at Physics Department, University of
California Riverside. Riverside, CA 92521-0413, U.S.A.}
}
\address{
$^a$Departamento de F\'{\i}sica.\\
Centro de Investigaci\'on y de Estudios Avanzados del IPN.\\
A.P. 14-740, C.P. 07000,  M\'exico, D.F., MEXICO.
}
\address{
$^b$Departamento de F\'{\i}sica Aplicada.\\
Centro de Investigaci\'on y de Estudios Avanzados del IPN.
Unidad M\'erida.\\ 
A.P. 73, Cordemex C.P. 97310, M\'erida, Yucat\'an, MEXICO.
}

\maketitle

\begin{abstract}
The existence of new quarks may lead to their mixing with ordinary quarks.
We are particularly interested in studying the consequences such mixings could
have upon the electromagnetic current.
We find that new strong flavor and parity violating electromagnetic
interactions may exist.
To be specific, for our analysis we work with the model 
$SU_2\otimes U_1\otimes \hat{SU}_2\otimes\hat U_1$,
which doubles the standard electroweak sector, both in gauge bosons and in 
quarks.
\end{abstract}

\paragraph*{\bf Introduction.}

It is well know that if new or exotic fermions exist their mixings with the
ordinary fermions will lead in general to flavor changing neutral current
interactions\cite{1}.
But as far as we know, it has not been discussed in detail if mixings of that
nature may lead to flavor and parity violating electromagnetic interactions.
At first glance, one might think that this is not the case, but a careful
analysis may lead to the opposite point of view, as we shall show in what
follows.
Put in rather reversed terms, one may ask what are the conditions for the
elimination of flavor and parity violating contributions to electromagnetic
current and if such conditions can always be met with certainty.
We shall see that, if those mixings exist then the latter contributions must
also exist.
The only way to eliminate them is to prohibit such mixings.

For the sake of definiteness, we shall work with the extension of the minimal 
electroweak model $SU_2\otimes U_1$ given\cite{2} by 
$SU_2\otimes U_1\otimes \hat{SU}_2\otimes\hat U_1$, which doubles the gauge 
boson sector, and with new so-called mirror quarks which double the quark
content.
We shall assume manifest left-right symmetry, which means that the gauge
constants of the $SU_2$'s and $U_1$'s obey $g=\hat g$ and $g'=\hat g'$,
respectively.
To keep the algebra simple we shall work with two families of ordinary quarks,
$u$, $d$, $s$, and $c$ and with two families of ``mirror'' counter-parts which
we will denote with hats, $\hat u$, $\hat d$, $\hat s$, and $\hat c$.

The higgs content frequently discussed consists of the doublet 
$\varphi(\frac{1}{2},Y_\varphi,0,0)$ and a mirror doublet
$\hat\varphi(0,0,\hat\frac{1}{2},\hat Y_{\hat\varphi})$.
These two leave massless one of the mirror neutral gauge bosons, the paraphoton.
We are interested specifically in the case in which only one massless neutral
gauge bosons exists and which couples to all the quarks present, i.e., the
mirror quarks will be assigned electric charges.
In these circumstances, the paraphoton should become a very massive neutral
boson.
To achieve this goal, new higgses must be introduced.
A bidoublet $\Phi(\frac{1}{2},Y_\Phi,\frac{1}{2}^*,-Y_\Phi)$ is a natural
choice.
Bisinglets $\Psi(0,Y_{\Psi},0,-Y_{\Psi})$ are also welcome.
The vacuum expectation values (VEV's) of all these higgses will be denoted by
$v$, $\hat v$, $k_1$, $k_2$ and $V$.
These VEV's will give masses to all the gauge bosons, but the photon.
A detailed analysis of this sector shows\cite{3} that in order to have two
light (the already observed ones) bosons, one charged and one neutral, it is
necessary that
$V^2\gg\hat v^2\gg v^2\sim k_1^2,k^2_2$
or that
$\hat v^2\gg V^2\gg v^2\sim k^2_1, k^2_2$. 
In both options the bisinglet is required to exist.

\paragraph*{\bf Quark masses and mixings.}

Our frame work is now set up. 
Let us next concentrate in the mixings of quarks.
The masses of the quarks will be generated by spontaneous symmetry-breaking
(SSB) at the several Yukawa quark-higgs couplings.
Straight-forward calculations lead to (a similar expression is found for $u$ and
$c$ type quarks) 

\begin{equation}
(\bar q)_L (m_{ij}) (q)_R + \mbox{\rm c.c.} = (\bar q)_L
\left(
\begin{array}{cc}
(g'_{ij}) & (\ell'_{ij}) \\
(j'_{ij})  & (\hat g'_{ij}) \\
\end{array}
\right)
(q)_R + \mbox{\rm c.c.},
\label{uno}
\end{equation}

\noindent
where, restricting ourselves to two families
$(q)_{L,R}=(d,s,\hat d,\hat s)_{L,R}$.
Thus $(m_{ij})$ is a 4$\times$4 and $(g'_{ij})$, 
$(\ell'_{ij})$, $(j'_{ij})$, and $(\hat g'_{ij})$ are 2$\times$2 matrices. 
The primes on the latter stand for the absorption of the VEV's into the 
corresponding Yukawa coupling constants.
These four last matrices are generated by the quark couplings with $\varphi$,
$\Phi$, $\Psi$, and $\hat\varphi$, respectively.
Left-right exchange symmetry, $\varphi\leftrightarrow\hat\varphi$,
$\Phi\leftrightarrow\Phi^\dagger$, $\Psi\leftrightarrow\Psi^\dagger$,
$q_L\leftrightarrow\hat q_R$, $q_R\leftrightarrow\hat q_L$, requires
$(g_{ij})=(\hat g_{ij})^\dagger$, $(j_{ij})=(j_{ij})^\dagger$, and 
$(\ell_{ij})=(\ell_{ij})^\dagger$.
We shall ignore CP violation and accordingly $(m_{ij})$ may be taken to be real.

The diagonalization of Eq.~(\ref{uno}) must go in two steps.
First we must extract the large angles and second the small ones.
Since one expects mirror matter to be much heavier than ordinary one, we must
have $(\hat g'_{ij})\gg (g'_{ij})$. 
We make a diagonalization of the block-diagonal part of $(m_{ij})$ containing
these two 2$\times$2 submatrices.
The diagonalizing matrix will also be block-diagonal and its effect upon
$(q)_{L,R}$ will be

\begin{equation}
(q^0)_{L,R} = 
\left(
\begin{array}{cc}
(\theta_{L,R}) & 0 \\
0 & (\hat\theta_{L,R})
\end{array}
\right)
(q)_{L,R},
\label{dos}
\end{equation}

\noindent
where $(\theta_{L,R})$ and $(\hat\theta_{L,R})$ are 2$\times$2 rotation matrix
and $\theta_{L,R}$ and $\hat\theta_{L,R}$ the corresponding rotation angles.
These angles will be large and the quarks obtained will the first approximation
to the quark mass eigenstates.
They can be assigned strong-flavors at this point, the index zero stands for
all this.
The initial mass matrix becomes 

\begin{equation}
(m_{ij})\quad{\longrightarrow\atop{\theta,\hat\theta}}\quad
\left(
\begin{array}{cc}
\begin{array}{cc}
m^0_d & 0 \\
0 & m^0_s
\end{array}
& (\Delta m^{(1)}_{ij}) \\
(\Delta m^{(2)}_{ij}) &
\begin{array}{cc}
\hat m^0_d & 0 \\
0 & \hat m^0_s
\end{array}
\end{array}
\right)
\label{tres}
\end{equation}

\noindent
where $\hat m^0_d$, $\hat m^0_s\gg m^0_d$, $m^0_s$.
The second step, the final rotation, will make (\ref{tres}) completely diagonal
and will yield the physical quarks.
In order that this rotation does not alter the above mass hierarchy, it is
necessary to have $(\hat g'_{ij})\gg (g'_{ij})\sim(\ell'_{ij})\sim(j'_{ij})$,
which in turn requires

\begin{equation}
\hat m^0_d,\ \hat m^0_s\ \gg\ m^0_d,\ m^0_s\ \sim \Delta m_{ij}^{(1)} \sim 
\Delta m_{ij}^{(2)}.
\label{cinco}
\end{equation}

\noindent
This means that the angles in the second rotation must be very small and they 
can be retained to first order.
The complete rotation matrices (in 2$\times$2 block form) are

\begin{equation}
R_{L,R} =
\left(
\begin{array}{cc}
I & (\epsilon^{L,R}_{ij}) \\
(-\epsilon^{L,R}_{ij}) & I
\end{array}
\right)
\left(
\begin{array}{cc}
(\theta_{L,R}) & 0 \\
0 & (\hat\theta_{L,R})
\end{array}
\right)
\label{seis}
\end{equation}

\noindent
where $\epsilon^{L,R}_{ij}\ll 1$.
The mass matrix is then diagonalized to 

\begin{equation}
R_L (m_{ij}) R^\dagger_R = 
\left(
\begin{array}{cc}
\begin{array}{cc}
m_d & \\
& m_s
\end{array}
& 0 \\
0 &
\begin{array}{cc}
\hat m_d & \\
& \hat m_s
\end{array}
\end{array}
\right)
\label{siete}
\end{equation}

\noindent
and $(q^0)_{L,R}$ are replaced by the physical quarks, $(q^{ph})_{L,R}$
which in turn become the new strong-flavor eigenstates. 
The small angles are determined in terms of the entries of (\ref{tres}).
We shall not display their detailed expression here, but it must be mentioned
that they are all of the order $m_{d,s}/\hat m_{d,s}$ and that it is necessary
$m_s-m_d$ not be small.

\paragraph*{\bf Quark anomalous magnetic moments and electromagnetic
interactions.}

The next subject to be discussed is the anomalous magnetic moments of quarks.
Although quarks are initially assumed to be point-like, QCD will dress 
them up, so to speak, and they will gain anomalous magnetic moments\cite{4}.
How to deal with these QCD-induced anomalous magnetic moments of quarks is the
central issue of the present analysis.
We shall exploit the fundamental property of the standard model: The
commutativity of the electroweak and QCD sectors. 
Our discussion will be by necessity qualitative, however we intend to keep it 
as general as possible.

In setting up an effective lagrangian, one usually (i) starts from the standard
model basic lagrangian, (ii) applies SSB, (iii) diagonalizes the mass matrices
and identifies the physical quarks, and finally (iv) uses QCD.
The explicit invariance of operators is lost after step (ii).
In contrast, we propose to follow the path (i) start from the basic lagrangian,
(ii) apply QCD, (iii) introduce SSB and, (iv) finally, diagonalize the mass
matrices and identify physical quarks.
We shall call the first path A and the second one B.
The advantage of B is that the effective operators induced by QCD must all
respect the electroweak gauge symmetry explicitly.

The appearance of QCD-induced anomalous magnetic moment operators of quarks 
can only take place after SSB, because prior to it one cannot write down 
electroweak invariant operators of this type. The reason for this is that $L$ 
and $R$ quarks must be connected to one another.
To achieve this connection it is necessary that the higgses intervene, only
then can invariant operators of magnetic moment type be written down.

Operators of this type will be induced by SSB upon graphs, for example, like 
$\varphi+q_L\to q_R+W^0$ or $\Phi+\hat q_L\to q_R+\hat W_3$.
SSB and the necessary rotations will yield operators corresponding to 
$q^{ph}_L\to q^{ph}_R+A_\mu$ or $\hat q^{ph}_L\to q^{ph}_R+A_\mu$, where
$A_\mu$ represents the photon field.
They are anomalous magnetic moment operators.
These latter operators are dressed by QCD and, as we have just discussed,
intimately related to the higgses.
Right after SSB, the anomalous magnetic moment 4$\times$4 matrix will look like
(in terms of 2$\times$2 matrices and suppressing Lorentz indices and
$\sigma_{\mu\nu}$ matrices)

\begin{equation}
(\bar q)_L (\mu_{ij}) (q)_R + \mbox{\rm c.c.}\ =\ (\bar q)_L 
\left(
\begin{array}{cc}
(\nu_{ij}) & (\nu'_{ij}) \\
(\nu''_{ij}) & (\hat\nu_{ij})
\end{array}
\right)
(q)_R + \mbox{\rm c.c.}
\label{ocho}
\end{equation}

\noindent
The rotations with the large angles of Eq. (\ref{dos}) will lead to 

\begin{equation}
(\mu_{ij})\quad{\longrightarrow\atop{\theta,\hat\theta}}\quad
\left(
\begin{array}{cc}
\begin{array}{cc}
\mu_d & 0 \\
0 &\mu_s
\end{array}
& (\Delta\gamma_{ij}) \\
(\Delta\gamma'_{ij}) &
\begin{array}{cc}
\hat\mu_d & 0 \\
0 & \hat\mu_s
\end{array}
\end{array}
\right).
\label{nueve}
\end{equation}

\noindent
Our experience with magnetic moments shows that their magnitudes are order of
magnitude inversely proportional to the masses for the particles.
So one must require that their hierarchy be

\begin{equation}
\mu_d,\ \mu_s\ \gg\ \hat\mu_d,\ \hat\mu_s\ \simeq\ \Delta\gamma_{ij}, 
\ \Delta\gamma_{ij'}.
\label{diez}
\end{equation}

The second rotation with the small angles will lead to the matrix (to first 
order in the small angles)

\begin{equation}
\left(
\begin{array}{cccc}
\mu_d &
\mu_d\epsilon^R_{21}+\mu_s\epsilon^L_{12} &
\mu_d\epsilon^R_{31}+\Delta\gamma_{11} &
\mu_d\epsilon^R_{41}+\Delta\gamma_{21} \\
\mu_d\epsilon^L_{21}+\mu_s\epsilon^R_{12} &
\mu_s &
\mu_s\epsilon^R_{32}+\Delta\gamma_{21} &
\mu_s\epsilon^R_{42}+\Delta\gamma_{22} \\
\mu_d\epsilon^L_{31}+\Delta\gamma'_{11} &
\mu_s\epsilon^L_{32}+\Delta\gamma'_{12} &
\hat\mu_d &
0 \\
\mu_d\epsilon^L_{41}+\Delta\gamma'_{21} &
\mu_s\epsilon^L_{42}+\Delta\gamma'_{22} &
0 &
\hat\mu_s
\end{array}
\right),
\label{once} 
\end{equation}

\noindent
there will be an analogous matrix for the complex conjugate anomalous magnetic
moments.
Matrix (\ref{once}) and its c.c.\ will be sandwiched between the physical
quarks and the electromagnetic field $A_\mu$ will appear in the
$F_{\mu\nu}$ stress tensor.
When all the terms are collected there will appear effective electromagnetic
interactions diagonal in the quark fields, like
$\mu_d\bar d^{ph}\sigma_{\mu\nu} d^{ph}F_{\mu\nu} + \mu_s \bar s^{ph} 
\sigma_{\mu\nu} s^{ph}F_{\mu\nu} + \mbox{(mirror part)}$.
These terms will be accompanied by non-diagonal operators, namely, 
$\bar d^{ph}\sigma_{\mu\nu}(g^{ds}_V + g^{ds}_A \gamma_5)s^{ph} +
\bar s^{ph}\sigma_{\mu\nu}(g^{sd}_V + g^{sd}_A \gamma_5)d^{ph} +
\mbox{(mixed ordinary-mirror terms)}$,
here there will not appear non-diagonal mirror-mirror terms.
The effective tensor and axial-tensor couplings are
$g^{ds}_V = g^{sd}_V = (\mu_s - \mu_d) (\epsilon^L_{12} + \epsilon^R_{12})$
and
$g^{ds}_A = -g^{sd}_A = (\mu_s + \mu_d) (\epsilon^L_{12} - \epsilon^R_{12})$.

From this analysis we may conclude that the existence of mirror matter and its 
mixing with ordinary matter will lead in general to the appearance of flavor
and parity violating terms in the electromagnetic interactions.

\paragraph*{\bf Discussion.}

Let us first discuss what condition must be met for the above flavor and 
parity violating terms in the electromagnetic current to disappear.
We have already assumed that the 2$\times$2 matrices of magnetic moments in the
diagonal of Eq.~(\ref{ocho}) are diagonalized into the 2$\times$2 diagonal
blocks of (\ref{nueve}), which is what one would ordinarily assume for the
minimal electroweak sector. 
If we now require that (\ref{once}) be diagonal too, then in its upper
2$\times$2 diagonal block we should have
$\mu_d\epsilon^R_{21} + \mu_s\epsilon^L_{12} = 0$
and
$\mu_d\epsilon^L_{21} + \mu_s\epsilon^R_{12} = 0$.
This is a system of two homogeneous equations for $\epsilon^R_{12}$ and 
$\epsilon^L_{12}$ (remember $\epsilon^R_{21} = -\epsilon^R_{12}$ and 
$\epsilon^L_{21} = -\epsilon^L_{12}$).
In order to have non-zero $\epsilon^R_{12}$ and $\epsilon^L_{12}$, the
determinant of the system must vanish.
This means that 

\begin{equation}
\mu^2_d = \mu^2_s.
\label{dieciseis}
\end{equation}

\noindent
Eq.~(\ref{dieciseis}) could only be satisfied in the strong-flavor symmetry
limit.
Since such symmetries are known to be broken (\ref{dieciseis}) cannot be
satisfied and, accordingly, (\ref{once}) remains non-diagonal.

One way out of this impasse is not to require that the 2$\times$2 blocks in the
diagonal of (\ref{ocho}) be diagonalized by the first large angle rotations. 
Detailed calculations show that then one can indeed impose that (\ref{once}) be 
completely diagonalized by the second small angle rotations.
This means that the price of eliminating electromagnetic flavor and parity
violations at the level
$SU_2\otimes U_1\otimes\hat{SU}_2\otimes\hat U$
is to already accept their existence at the level of the minimal
$SU_2\otimes U_1$.
This is a rather contradictory point of view.
It also means that the limit of the large gauge group (by sending the VEV's of
$\Phi$, $\Psi$, and $\hat\varphi$ to infinity) into the minimal one is far
from smooth.
This would render any attempt to estimate the QCD-induced anomalous magnetic
moments of ordinary quarks at level of $SU_2\otimes U_1$ meaningless\cite{4,5}.
One would face a new fine-tuning problem which would cast serious doubts on our
present understanding of the standard model.
This option is not satisfactory either. 
Another way out would be that (\ref{nueve}) be diagonalized very much as
(\ref{tres}) was diagonalized into (\ref{siete}).
For this to be the case the hierarchy (\ref{diez}) of the magnetic moment
matrix elements must be abandoned and, contrastingly, should be required to be
analogous to the hierarchy (\ref{cinco}) of mass matrix elements. 
This would lead to the very massive mirror quarks to have enormous anomalous 
magnetic moments, i.e., a very unphysical situation\cite{5}.

One is finally led to the choice either ordinary and mirror quarks do not mix at
all and electromagnetic interactions are always flavor and parity conserving or
they do mix and then non-vanishing flavor and parity violating electromagnetic
interactions appear.

The above analysis shows that the non-existence of flavor and parity violation
in electromagnetic interactions is not a question of fundamental principles,
but is simply a question of assumption.
The possibility of flavor and parity violations in electromagnetic interactions
is really an open question, which can be decided upon only by experiment.
All we can say as of now is that the $\epsilon^R_{ij}$ and $\epsilon^L_{ij}$
angles must be very small, in order to comply with the observed intensity of
parity and flavor violation in nature.

Our analysis has been qualitative in nature.
Nevertheless, we hope it has remained general enough to show that new flavor
and parity violating interactions may exist in nature and that they should be
considered seriously.

\paragraph*{\bf Acknowledgments.}

One the authors (A.G.) wishes acknowledge interesting discussions with G.~Kane
and P.~Langacker.
This work is supported in part by CONACyT (M\'exico).


\begin{references}
\bibitem{1}
P. Langacker and D. London, Phys. Rev. {\bf D38}, 886 (1988); and 
references therein.
\bibitem{2}
S. M. Barr, D. Chang, and G. Senjanovi\'c, Phys. Rev. Lett. {\bf 67}, 2765
(1992).
Here can be found references to prior models.
\bibitem{3}
O. Miranda, Ph.D. Thesis (CINVESTAV, 1997) and to be submitted for publication.
\bibitem{4}
L. Brekke, Ann. Phys. {\bf 240}, 400 (1995).
\bibitem{5}
H. Georgi, L. Kaplan, D. Morin, and A. Schenk, Phys. Rev. {\bf D51}, 3888
(1995).
\end{references}
\end{document}